\documentstyle[12pt] {article}
\def\beginrev{\begingroup}
\def\reviewer{\par{\bf Reviewer: }}

\def\address{\par{\bf Address: }\par
  \begingroup\leftskip=16pt\parskip=0pt\obeylines}
\def\author{\endgroup\par{\bf Author: }}
\def\shorttitle{\par{\bf Short title: }}
\def\journal{\par{\bf Journal: }} 
\def\mcno{\par{\bf Control number: }}
\def\mmrno{}
\def\rpclass{\par{\bf Primary classification: }}
\def\rsclass{\par{\bf Secondary classification(s): }}

\begin{document}

\begin{flushright}
To be published in \\
Mathematical Reviews
\end{flushright}

\beginrev 
\reviewer Luca Salasnich 
\address
Dipartimento di Matematica Pura ed Applicata
Universit\`a di Padova 
Via Belzoni 7
I-35131 Padova, Italy
\author C. Jung and T.H. Seligman
\shorttitle Integrability of the S--Matrix 
\journal Phys. Rep. {\bf 285}, 77--141 (1997)
\mcno 1 452 542 
\mmrno 
\rpclass 81Q50
\rsclass 70K50 
\vskip 0.5 truecm
\par
This interesting paper deals with the connection 
between the integrability of the 
scattering matrix $S$ and the integrability of the Hamiltonian $H$ 
for classical and quantum Hamiltonian systems. 
\par
The authors study the dynamics of a system described 
by the Hamiltonian $H$ and for which 
the asymptotic dynamics is given by the free 
Hamiltonian $H_0$. From these Hamiltonians they derive 
the quantum evolution operators $U(t)$ and $U_0(t)$, and also 
the corresponding classical flow maps, which are denoted by 
$\Phi (t)$ and $\Phi_0(t)$, respectively. Since 
the quantum $S$--operator is defined as 
$$
S= \lim_{T\to \infty}U_0(T)U(2T)U_0(-T) \; 
$$ 
[see, for instance, J.R. Taylor, {\it Scattering 
Theory}, Wiley, New York, 1972], they define a corresponding classical map 
$$
{\tilde M}=\lim_{T\to \infty}\Phi_0(T)\Phi (2T)\Phi_0(-T) \; .
$$ 
Then the authors introduce the classical map $M$, which is the map 
${\tilde M}$ reduced to the channel space and for fixed energy. 
For a system with $n$ degrees of freedom, 
where $M$ has a $(2n-2)$--dimensional domain, 
they call $M$ completely integrable if there are $n-1$ independent functions 
in involution such that their common level sets are invariant under $M$. 
It is shown that integrability of $M$ always leads to integrability 
of $H$, but that integrability of $H$ only leads to integrability of $M$ 
under an additional condition, namely the asymptotic forward and 
backward limits must coincide. 
\par
In the second part of the paper 
the authors transfer their classical results to quantum dynamics. 
For a system with $n$ degrees of freedom they call $S$ completely 
integrable if there are $n$ independent operators that commute among 
themselves and with $S$. Because in quantum mechanics the question of 
independence of a set of commuting operators is more subtle than 
in classical mechanics, they consider only operators which have a classical 
counterpart. In this way it is shown that 
integrability of $S$ always implies integrability 
of $H$, but integrability of $H$ implies integrability of $S$ 
under the same additional condition obtained in classical mechanics. 
\par
Finally the authors give some numerical example of breaking of 
integrability for the $S$--matrix. They show that the nearest neighbour 
spacing distribution $P(s)$ of the eigenphases of the $S$--matrix 
follows the Poisson distribution 
$$
P(s)=exp{(-s)} \; ,
$$ 
if the matrix is integrable,  and that $P(s)$ is very closed to the 
Wigner surmise 
$$
P(s)=(\pi s/2)\exp{(-\pi s^2/2)} \; ,
$$ 
if the classical map $M$ is dominated by chaos. 

\end{document}